\documentclass[manuscript]{emulateapj}

\usepackage{natbib}
\usepackage{color}
\usepackage{amsmath}
\usepackage{cases}
\usepackage{longtable}

\usepackage{footmisc} % use same footnote for multiple terms

\shorttitle{Magnetic outflows of NGC/,4388}
\shortauthors{A. Damas-Segovia et al.}

\slugcomment{}

\shorttitle{CHANG-ES VII:\\ Magnetic outflows from the Virgo cluster galaxy NGC\,4388}
\shortauthors{Damas-Segovia et al.}

\newcommand{\HI}{{\rm H\,\scriptstyle I}}

\newcommand{\OIII}{{\rm O\,\scriptstyle III}}
\newcommand{\Ha}{{\rm H\alpha}}
\newcommand{\kms}{\,{\rm km\,s^{-1}}}
\newcommand{\ccm}{\,{\rm cm^{-3}}}
\newcommand{\radm}{\,{\rm rad\,m^{-2}}}

\begin{document}


\title{CHANG-ES VII:\\ Magnetic outflows from the Virgo cluster galaxy NGC\,4388}


\author{A. Damas-Segovia$^1$, 
        R. Beck$^1$, 
        B. Vollmer$^2$, 
        T. Wiegert$^3$, 
        M. Krause$^1$, 
        J. Irwin$^3$,
        M. We{\.z}gowiec$^4$,
        J. Li$^5$,
        R-J. Dettmar$^6$,
        J. English$^7$,
        Q. D. Wang$^8$}
\affil{$^1$MPI f\"ur Radioastronomie, Auf dem H\"ugel 69, 53121 Bonn, Germany \\
$^2$Observatoire astronomique de Strasbourg, Universit\'e de Strasbourg, CNRS, UMR 7550, 11 rue de l'Universit\'e, \\ 67000 Strasbourg, France\\
$^3$Dept.   of  Physics,  Engineering  physics  \&  Astronomy, Queen's  University,  Kingston,  ON,  Canada,  K7L  3N6\\
$^4$Obserwatorium Astronomiczne Uniwersytetu Jagiello{\'n}skiego, ul. Orla 171, 30-244 Krak{\'o}w, Poland\\
$^5$Dept. of Astronomy, University of Michigan, 311 West Hall, 1085 S. University Ave., Ann Arbor, MI 48109, USA\\
$^6$Astronomisches Institut, Ruhr-Universit\"{a}t Bochum, 44780 Bochum, Germany\\
$^7$Department of Physics and Astronomy, University of Manitoba, Winnipeg, Manitoba, R3T 2N2, Canada\\
$^8$Dept. of Astronomy, University of Massachusetts, 710 North Pleasant St., Amherst, MA, 01003, USA}

\email{adamas@mpifr-bonn.mpg.de}


%\altaffiltext{1}{Visiting Astronomer, Cerro Tololo Inter-American Observatory.
%CTIO is operated by AURA, Inc.\ under contract to the National Science
%\altaffiltext{2}{Society of Fellows, Harvard University.}
%\altaffiltext{3}{present address: Center for Astrophysics,
%    60 Garden Street, Cambridge, MA 02138}
%\altaffiltext{4}{Visiting Programmer, Space Telescope Science Institute}
%\altaffiltext{5}{Patron, Alonso's Bar and Grill}


\begin{abstract}

We investigate the effects of ram pressure on the ordered magnetic field of a galaxy hosting a radio halo and strong nuclear outflows.
New radio images in total and polarized intensity of the edge-on Virgo galaxy NGC\,4388 were obtained within the CHANG-ES EVLA project.
The unprecedented noise level reached allows us to detect striking new features of the ordered magnetic field.
The nuclear outflow extends far into the halo to about 5\,kpc from the center and is spatially correlated with the $\Ha$ and X-ray emission. For the first time, the southern outflow is detected. 
Above and below both spiral arms we find extended blobs of polarized emission with an ordered field oriented perpendicular to the disk. The synchrotron lifetime of the cosmic ray electrons (CREs) in these regions yields a mean outflow velocity of $(270\pm70)\kms$, in agreement with a galactic wind scenario. The observed symmetry of the polarized halo features in NGC 4388 excludes a compression of the
halo gas by the ram pressure of the intra-cluster medium (ICM). The assumption of equilibrium between the halo pressure and the ICM ram pressure yields
an estimate of the ICM density that is consistent with both the ICM density derived from X-ray observations and the recent \textit{Planck}
Sunyaev-Zel'dovich measurements. 
%Allowing for ICM clumping might lower the actual ICM ram
%pressure by up to a factor of two compared to its value based on a
%smooth continuous ICM distribution. 
The detection of a faint radio halo around cluster galaxies could thus be used for an estimate of ICM ram pressure.

\end{abstract}


\keywords{galaxies: individual: NGC\,4388 -- galaxies: clusters: individual: Virgo Cluster -- galaxies: halos -- galaxies: intergalactic medium -- galaxies: jets -- galaxies: magnetic fields}



\section{Introduction}

NGC\,4388 is an almost edge-on SA(s)b galaxy (inclination 79\degr) in the Virgo cluster (at about $17\,\rm{Mpc}$
distance), located about $1.3\degr$ ($\approx400\,\rm{kpc}$) west from the center of the cluster \citep{Chung}.
The interstellar medium (ISM) of NGC\,4388 has undergone a stripping event by ram pressure, evident from the rapid decline of star
formation $(190\pm30)\,\rm{Myr}$ ago \citep{Pappalardo}. Like many other galaxies in cluster environments, NGC\,4388 is a $\HI$-deficient galaxy that lost about 85\% of its $\HI$ mass \citep{Cayatte}. The $\HI$ disk is strongly
truncated within the optical disk. An $\HI$ plume extends up to 100\,kpc out from the galaxy plane \citep{Oosterloo}. The gas stripping is possibly the result of the interaction between the
galaxy and the ICM. \cite{Vollmer_2003} estimated that the galaxy passed close to the
cluster center about $120\,\rm{Myr}$ ago. A more recent model yields a timescale of about $200\,\rm{Myr}$ (Vollmer, priv. comm.).

Early VLA studies of this galaxy revealed a bright double source in the nucleus and an outflow lobe that
opens up like an hourglass and extends to about 1.5\,kpc from the center \citep{Hummel_1983,Hummel,Kukula,Falcke}.
Observations with very-long-baseline interferometry (VLBI) revealed a radio jet with $\cong0.5$\,pc extent \citep{Giroletti}. A circumnuclear
disk of $\cong0.1$\,pc radius, oriented edge-on and almost parallel to the galaxy plane, was detected by its
water maser emission \citep{Kuo}. This indicates that the jet and the outflow emerge almost perpendicular to the
galaxy plane.

Previous VLA and Effelsberg radio continuum observations showed an extended halo and extended features out of the disk \citep{Vollmer_2010,Wezgowiec}. These features may be related to the outflow event seen in $\Ha$ and $\OIII$
observations \citep{Yoshida_2002,Yoshida_2004}.

These deep optical spectroscopic observations are crucial to understanding the complexity
and the dynamics of the outflow of ionized gas from this galaxy. \cite{Yoshida_2002} concluded that the extended
emission-line region is gas stripped by the action of ram pressure and photo-ionized by the radiation
from the AGN.

High resolution X-ray observations with \textit{Chandra X-ray Observatory} show emission from the outflow as well as from the radio jet \citep{Iwasawa}. On the other hand, lower resolution X-ray observations with XMM-Newton reveal the hot halo surrounding this galaxy. This halo might be shaped by the Mach cone created by the supersonic speed of the galaxy in that medium \citep{Wezgowiec_Mach}. The clear correlation between features in optical, radio and X-ray frequencies suggests a common origin.

%%%%%%%%%%%%%%%%%%%%%%% NGC 4388 properties %%%%%%%%%%%%%%%%%%%%%%%%%%%%%%%%%%%%%%%%%%%%%%%%%%%%%%%%%%%%%%%%%%
%%%%%%%%%%%%%%%%%%%%%%%%%%%%%%%%%%%%%%%%%%%%%%%%%%%%%%%%%%%%%%%%%%%%%%%%%%%%%%%%%%%%%%%%%%%%%%%%%%%%%%%%%%%%%%%%%%%%%%%%%
%%%%%%%%%%%%%%%%%%%%%%%%%%%%%%%%%%%%%%%%%%%%%%%%%%%%%%%%%%%%%%%%%%%%%%%%%%%%%%%%%%%%%%%%%%%%%%%%%%%%%%%%%%%%%%%%%%%%%%%%%

\begin{table}
\caption{NGC\,4388 properties}              % title of Table
\centering                                      % used for centering table
\begin{tabular}{l l}          % centered columns (4 columns)
\hline\hline                        % inserts double horizontal lines
% table heading                               % inserts single horizontal line
R.A. (J2000) &$\rm{12^h25^m46.75^s}$\\
Dec. (J2000) &$\rm{12^d39^m43.5^s}$\\
Type & SA(s)b \\
Inclination\footnote{Inclination, $i = 3\,\rm{deg} + cos^{-1}(\sqrt{((b/a)^2 - 0.2^2)/(1 - 0.2^2))}$, where b/a is the minor to major axis ratio.} & $79\degr$ \\
Position angle & $92\degr$ \\
$d_{25}$\footnote{Observed blue diameter at the 25th mag $\rm{arcsec}^{-2}$ isophote.}& $5.6'$\\
Distance & $17\,\rm{Mpc}$\\
\hline
\end{tabular}
\label{table:obs}
\end{table}

%%%%%%%%%%%%%%%%%%%%%%%%%%%%%%%%%%%%%%%%%%%%%%%%%%%%%%%%%%%%%%%%%%%%%%%%%%%%%%%%%%%%%%%%%%%%%%%%%%%%%%%%%%%%%%%%%%%%%%%%%
%%%%%%%%%%%%%%%%%%%%%%%%%%%%%%%%%%%%%%%%%%%%%%%%%%%%%%%%%%%%%%%%%%%%%%%%%%%%%%%%%%%%%%%%%%%%%%%%%%%%%%%%%%%%%%%%%%%%%%%%%
%%%%%%%%%%%%%%%%%%%%%%%%%%%%%%%%%%%%%%%%%%%%%%%%%%%%%%%%%%%%%%%%%%%%%%%%%%%%%%%%%%%%%%%%%%%%%%%%%%%%%%%%%%%%%%%%%%%%%%%%%

The distribution of polarized radio emission in a cluster galaxy tends to be asymmetric as a consequence of
interaction with the ICM \citep{Vollmer_2007,Vollmer_2013}.
This interaction enhances the polarized flux on the side which is facing the ram pressure due to the motion
of the galactic disk in the denser environment. The $\HI$ plume detected by \cite{Oosterloo} represents clear evidence of interaction between the galaxy and the ICM of Virgo toward the southern part of its disk.

Little is known so far about the role of magnetic fields in the outflows of NGC\,4388. Polarized radio emission is a
signature of ordered magnetic fields. Previous low-resolution Effelsberg observations by \citet{Wezgowiec} showed a
large-scale magnetic field that is inclined with respect to the disk plane. Higher-resolution VLA observations by \citet{Vollmer_2010} show a complex magnetic field structure which is different in the
outflow, the disk and the halo.

\cite{Vollmer_2009} performed simulations with a sticky particle code \citep{Vollmer_2001} to quantify the outflow
event that takes place in NGC\,4388. They were able to reproduce with high fidelity the direction, speed and density
of the stripped gas. An important part of that study was to observe the evolution of the ram pressure of the ICM as a function
of time. In that model the predicted ram pressure at the current stage of the galaxy's orbit is in a decreasing phase. They estimated a ram pressure of $1\times 10^{-11}\,\rm{dyn\,cm}^{-2}$ at the present time. Within this scenario, it is expected that the action of ram pressure would transform the morphology of the gaseous galactic halo. The side of the galaxy facing ram pressure would be compressed leading to a sharp smooth edge of the radio continuum emission, with no extensions toward the outskirts of the halo.

We intend to investigate the impact of ram pressure on the magnetic field of NGC\,4388 with help of the new
radio polarization observations. In particular, wide-band polarimetric studies are of crucial importance for this
purpose, since they reach lower limits of rms and therefore allow us to detect faint structures out in the halo of a galaxy where a weaker magnetic field is expected. In the current study we make use of improved instruments and new techniques. Higher resolution
and sensitivity reveal many details of the magnetic field structure of NGC\,4388 that were never
observed before. New features of the galactic magnetic field in NGC\,4388 extending up to 5\,kpc out in the halo on both sides of the galactic disk are revealed. They are in conflict with models that include a continuous smooth ICM density profile which predict an asymmetric gas distribution within the galactic halo due to ISM compression.

In Sect.~\ref{observations} we present the observations and the data reduction. Images of radio total power and polarized emission as well as an analysis of the magnetic field of NGC\,4388 can be found in Sect.~\ref{results}.
A discussion on the different features seen in these new observations follows in Sect.~\ref{Interpretation}. We suggest in Sect.~\ref{Ram pressure} that ICM clumpiness as a possible explanation for the discrepancy between the ram pressure derived from observations and predicted by dynamical models. Sect.~\ref{summary} contains a summary of the main points of the present study.

%________________________________________________________________

%__________________________________________________________________

%%%%%%%%%%%%%%%%%%%%%%% Table of Observations %%%%%%%%%%%%%%%%%%%%%%%%%%%%%%%%%%%%%%%%%%%%%%%%%%%%%%%%%%%%%%%%%%
%%%%%%%%%%%%%%%%%%%%%%%%%%%%%%%%%%%%%%%%%%%%%%%%%%%%%%%%%%%%%%%%%%%%%%%%%%%%%%%%%%%%%%%%%%%%%%%%%%%%%%%%%%%%%%%%%%%%%%%%%
%%%%%%%%%%%%%%%%%%%%%%%%%%%%%%%%%%%%%%%%%%%%%%%%%%%%%%%%%%%%%%%%%%%%%%%%%%%%%%%%%%%%%%%%%%%%%%%%%%%%%%%%%%%%%%%%%%%%%%%%%

\begin{table}
\caption{Details of VLA observations}              % title of Table
\centering                                      % used for centering table
\begin{tabular}{l l}          % centered columns (4 columns)
\hline\hline                        % inserts double horizontal lines
% table heading                               % inserts single horizontal line
Dates of observations & 19 Dec 2011 and 08 Apr 2012 \\
Configurations & C, D\\
Central frequency & 6.0\,GHz \\
Bandwidth & 2.0\,GHz \\
Spectral channels per window & 64  \\
Channel width & 2\,MHz  \\
Spectral windows & 16  \\
Primary calibrator & 3C286 \\
Secondary calibrator & J1254+1141 \\
Polarization leakage calibrator & J1407+2827 \\
\hline
\end{tabular}
\label{table:obs}
\end{table}

%%%%%%%%%%%%%%%%%%%%%%%%%%%%%%%%%%%%%%%%%%%%%%%%%%%%%%%%%%%%%%%%%%%%%%%%%%%%%%%%%%%%%%%%%%%%%%%%%%%%%%%%%%%%%%%%%%%%%%%%%
%%%%%%%%%%%%%%%%%%%%%%%%%%%%%%%%%%%%%%%%%%%%%%%%%%%%%%%%%%%%%%%%%%%%%%%%%%%%%%%%%%%%%%%%%%%%%%%%%%%%%%%%%%%%%%%%%%%%%%%%%
%%%%%%%%%%%%%%%%%%%%%%%%%%%%%%%%%%%%%%%%%%%%%%%%%%%%%%%%%%%%%%%%%%%%%%%%%%%%%%%%%%%%%%%%%%%%%%%%%%%%%%%%%%%%%%%%%%%%%%%%%

\begin{figure*}[t!]
\centering
\includegraphics[width=0.8\textwidth,trim = 0mm 20mm 0mm 10mm,clip]{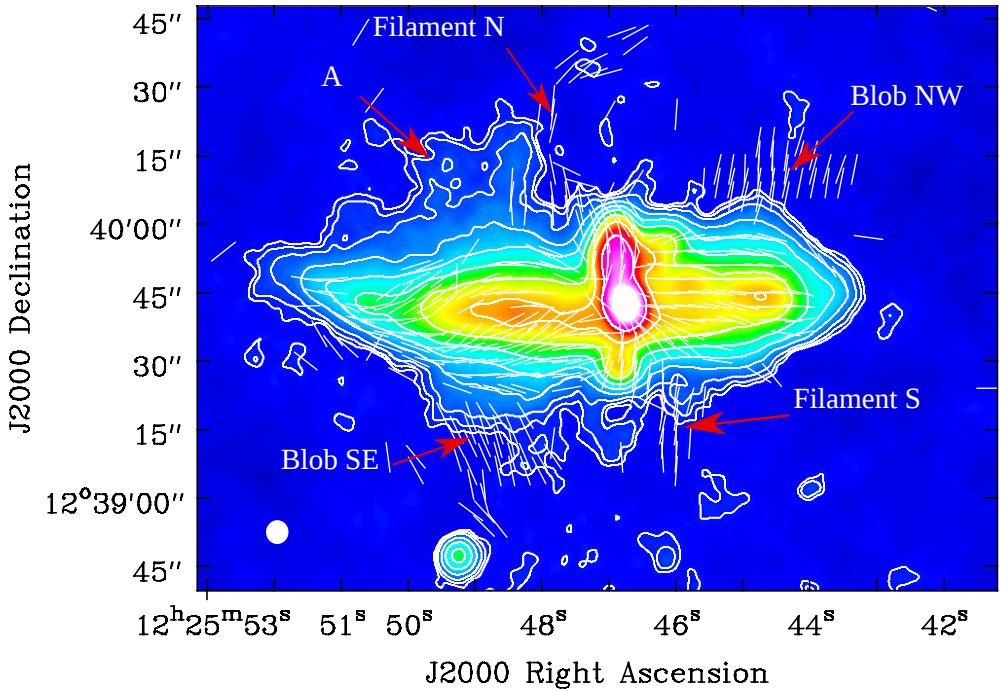}
\caption{\small{Total intensity at 6\,GHz in contours plus magnetic vectors with the same vector length, corrected for Faraday rotation, with a resolution of $4.99\arcsec\,\times\,4.69\arcsec$ and an rms noise of $3.5\,\mu \rm{Jy/beam}$.
The contour levels are $(3, 4, 6, 12, 24, 48, 96, 200, 500)\,\times 3.5\,\mu\rm{Jy/beam}$.
A combination of the data from the C and D configurations of the VLA was used to make the total intensity map
whereas only C configuration data was necessary to create the map of magnetic vectors.
Both maps have been cleaned with a robust 2 weighting. The galaxy moves in the SW direction.}}
\label{TotalI_PA}
\end{figure*}

\begin{figure}[t!]
\includegraphics[width=1\columnwidth,trim = 0mm 20mm 0mm 10mm,clip]{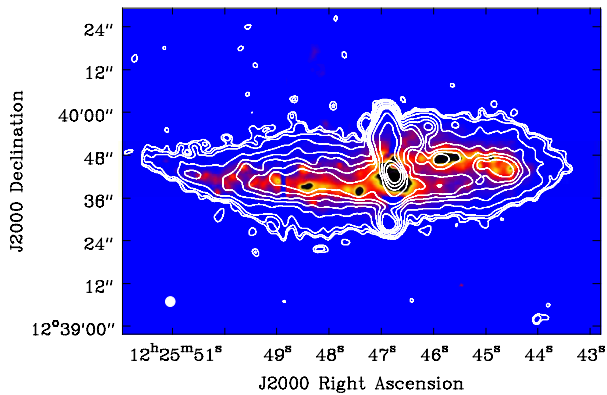}
\caption{\small{Total intensity contours plotted over $\Ha$ map from \cite{Yoshida_2002}.
Contours are $(3, 4, 6, 12, 24, 36, 48, 64, 128, 256, 512, 1024)\,\times 3.3\,\mu\rm{Jy/beam}$. This map was made with data of C configuration with robust 0 weighting. The resolution of the radio total intensity is $2.76\arcsec\,\times\,2.67\arcsec$ and the rms noise is $3.3\,\mu \rm{Jy/beam}$.}}
\label{spiralArms}
\end{figure}

%%%%%%%%%%%%%%%%%%%%%%

\begin{figure*}[t!]
\centering
\includegraphics[width=0.8\textwidth,trim = 0mm 20mm 0mm 10mm,clip]{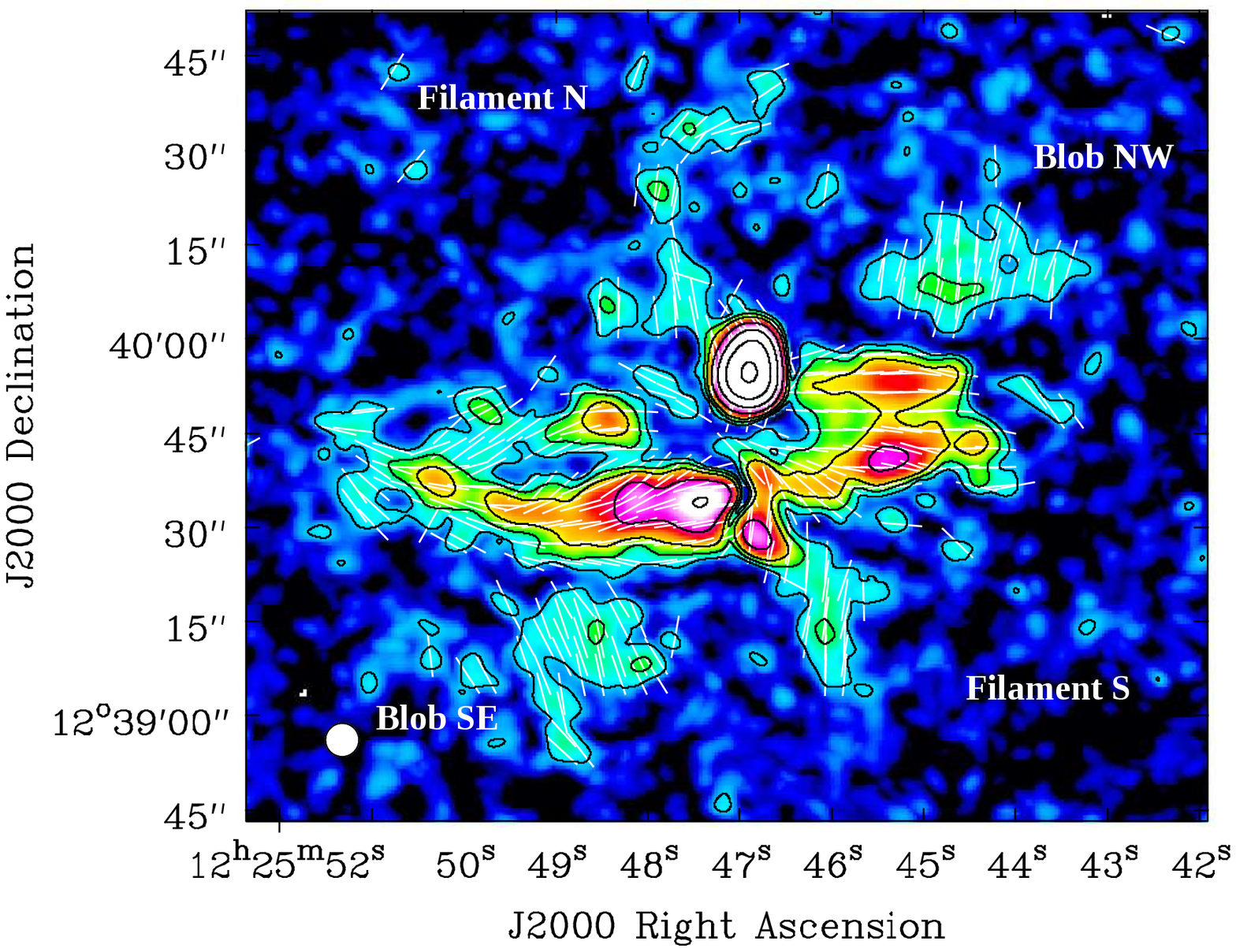}
\caption{\small{Linearly polarized intensity at 6\,GHz in contours plus magnetic vectors with the same vector length, corrected for Faraday rotation, with a resolution of $5.33\arcsec\,\times\,5.33\arcsec$ and an rms noise of $2.3\,\mu \rm{Jy/beam}$. The contour levels are $(3, 5, 8, 16, 32, 64, 128)\,\times 2.3\,\mu\rm{Jy/beam}$. Both maps have been cleaned with a robust 2 weighting.}}
\label{PI_PA}
\end{figure*}

\section{Observations and data reduction}
\label{observations}
As a part of the CHANG-ES project (Continuum Halos in Nearby Galaxies - an EVLA Survey; \citealt{Irwin}), the
data set presented here was observed in C band (5--7\,GHz) with the the Karl G. Jansky Very Large Array (VLA) in D configuration \citep{Wiegert_2015} and C configuration (this paper).
Details of the observations are given in Table~\ref{table:obs}.
The large bandwidth at this frequency gives a high sensitivity never reached before in imaging of galaxies and allows the application of
the RM (rotation measure) Synthesis technique \citep{Brentjens,Heald_2009,Heald_2015} that increases the signal to noise ratio of the polarized intensity images compared to previous studies.

By combining the datasets of both array configurations we improved
the uv coverage and the signal-to-noise ratio substantially.

L-band data (1--2 GHz) were also obtained as part of the CHANG-ES project, but unfortunately the data sets in
this frequency band reveal large rms noise values for the NGC\,4388 field. This is due to the fact that
this galaxy is very close to the Virgo cluster core in the plane of the sky ($\approx\,1\degr$) and therefore
cleaning residuals from M\,87, which is the central source of this cluster, increase the noise. M\,84, a radio galaxy which lies right at the edge of the primary beam at this frequency range, also increases the noise of these observations. In addition, no polarization was found in the L band datasets probably due to a stronger depolarization effect by Faraday dispersion at these frequencies and higher noise.

The data reduction was done using the Common Astronomy Software Applications package (CASA\footnote{http://casa.nrao.edu/}). Radio frequency interference (RFI) flagging and calibration was performed for each individual channel. Imaging in Stokes I was done averaging the entire band of the dataset while imaging in Q and U was done for each of the 16 spectral windows to be combined into a cube. Natural weighting (robust $= 2$) was used during the cleaning process to ensure a higher sensitivity to extended emission.
Due to the good quality data only two runs of selfcalibration were made, one in phase and a second one in amplitude and phase.

For the first time for this galaxy, the polarization products of this dataset were obtained by applying the RM Synthesis method
\citep{Brentjens} to the 15 images of the usable spectral windows of $128\,\rm{MHz}$ each in Stokes Q and U.
One window was removed due to low signal-to-noise ratios.
Each individual spectral window image was smoothed to the resolution of the spectral window with the lowest
resolution and corrected for primary beam attenuation.
RM Synthesis allows us to recover most of the polarized emission that would be lost due to bandwidth depolarization
while averaging Q and U intensities over the whole bandwidth. The final polarized intensity map was obtained from the Q and U values at the maximum intensity peaks in the Faraday spectrum at each pixel in the sky plane. The bias generated in polarized intensity was then subtracted.
Thanks to this technique we reach a noise of $2.3\,\mu\rm{Jy/beam}$ at a resolution of $5.33\arcsec\,\times\,5.33\arcsec$ with an observation of 180 minutes on the target source (before flagging). To our knowledge, this is
the lowest rms for a polarization map yet accomplished for an external galaxy.

The wavelength coverage of our observations in the given wavelength range ($\Delta\lambda^2$) gives a resolution in Faraday space of
$\delta \phi \cong 2000\radm$, which is sufficient to distinguish components in the Faraday spectrum that
are separated by more than $\delta \phi/(2 S/N)$, where $S/N$ is the signal-to-noise ratio of the polarized
intensity. As Faraday depth of a few $100\radm$ are not expected from emitting regions along the line
of sight through galaxies, we presume that we detect only one Faraday component at all locations.

\section{Results}
\label{results}
\subsection{Images in total and polarized intensity}

The total intensity image obtained by combining D and C array observations from CHANG-ES
show new details in the morphology of this galaxy (see Fig.~\ref{TotalI_PA}). The prominent jet-like structure just north of the nucleus is clearly
detected, as in previous works \citep{Hummel,Kukula,Falcke}, but now the disk is seen with higher resolution and sensitivity. The high surface brightness disk ($\approx 700\,\mu \rm{Jy/beam}$) of these new observations is quite symmetric including two spiral arms (the northwestern arm being closer to us) observed in $\Ha$ (\cite{Yoshida_2002,Yoshida_2004}; Fig.\ref{spiralArms}). At intermediate surface brightness ($\approx 40\,\mu \rm{Jy/beam}$) we observe a pointed tail to the east (on the far east end of the galaxy) which bends toward the north. The southern part of that tail shows a sharp edge. A low surface brightness ($\approx 10\,\mu \rm{Jy/beam}$) diffuse
halo is also detected. The halo is more extended and patchy in the eastern side of the galaxy. In the northeastern region (see region marked as A in Fig.~\ref{TotalI_PA}) we find a similar structure between the total intensity radio emission and the $\Ha$ outflow.

Figure~\ref{PI_PA} shows the polarized
emission of NGC\,4388. The total power jet-like structure is also prominent in polarization. We detect strong polarized emission within the total power disk. The polarized intensity is strongest between the spiral arms as commonly observed in spiral galaxies \citep{Beck_2013}.
The rms noise in the polarized maps is lower than in the total intensity maps, therefore we are able to
see many details in polarization that are not revealed in total power.
Extended new features in polarized emission are discovered far away from the disk. Two almost vertical filaments have projected extensions of 50\arcsec\ ($\cong4.1$\,kpc) to the north and 20\arcsec\ ($\cong1.7$\,kpc) to the south.
Other intriguing features of these new observations are the two extended extra-planar regions of polarized emission at large distances ($\approx 3.7\,\rm{kpc}$) northwest and southeast from the galaxy center. In the following we will call these features the northwestern and southeastern blobs. They reveal an ordered magnetic field and CREs about $5\,\rm{kpc}$ projected distance from
the plane of the galaxy.

\subsection{The magnetic field in NGC\,4388}
\label{equi}

Figure~\ref{TotalI_PA} shows magnetic vectors in the disk parallel to the major axis as expected for edge-on galaxies \citep{Krause_2011}. In the southeastern part of the disk the polarization vectors are no longer parallel to the disk and run along the sharp edge of the total power emission. Vertical magnetic vectors can be seen in the central region along the jet-like structure. All magnetic field vectors outside the high and intermediate surface brightness of total power emission are almost vertical with respect to the galactic disk.

Using the revised equipartition formula by \citet{Beck_2005}, we calculated the ordered and total magnetic field strengths from the polarized and total intensities, respectively.
We assumed different path lengths and synchrotron spectral indices for the various parts of the galaxy (see
Table~\ref{B_all_parts_table}). The pathlengths through the emitting medium along the line of sight
are taken to be the same as the widths of the features in the sky plane.
The ratio between the number densities of cosmic ray protons and electrons in the relevant energy range of a few GeV
is assumed to be $\rm{K} = 100$. This value may be larger in the halo due to energy losses of CREs (see Sect.~\ref{Blobs}). 

We discuss the possibility of a larger proton-to-electron ratio K than the value
of 100 assumed to derive the values in Table~\ref{B_all_parts_table}. The total equipartition magnetic field $\rm{B_{eq}}$ depends on K as follows \citep{Beck_2005}:

\begin{equation}
B_{\rm{eq}}\propto K_{\rm{eq}}^{1/(3+\alpha)}\, ,
\end{equation}
where $B_{\rm{eq}}$ and $K_{\rm{eq}}$ are the total magnetic field and the proton to electron ratio from equipartition, respectively. 
In the absence of a cosmic ray source in the halo, we expect a steep spectral index ($\alpha > 1$ assuming $S_{\nu} \propto \nu^{-\alpha}$). Therefore, magnetic strength depends on $K_{\rm{eq}}$ in the following way: $B \propto K_{\rm{eq}}^{1/4 \, ...\, 1/5}$. Thus, the magnetic field strength is not very sensitive to $K_{\rm{eq}}$. An uncertainty in $K_{\rm{eq}}$ of a factor of 2 leads to an error in $B$ of about $20\%$.

\begin{figure}[t!]
\includegraphics[width=1\columnwidth,trim = 0mm 10mm 0mm 10mm,clip]{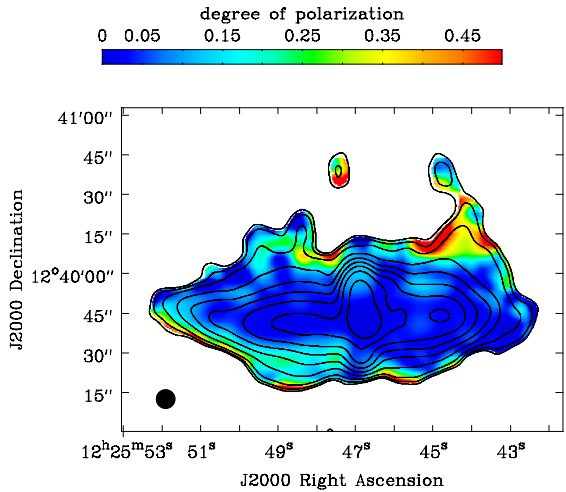}
\caption{\small{Contours of total intensity emission plotted with degree of polarization in
colour scale. Contours of total intensity emission are $(3, 4, 6, 12, 24, 48, 96, 200, 500)\,\times 6\,\mu\rm{Jy/beam}$.
The resolution is $7\arcsec\,\times\,7\arcsec$ and the rms noise is $6\,\mu\rm{Jy/beam}$.}}
\label{degree_pol}
\end{figure}

We also considered different spectral index values for the various parts of the galaxy. Large deviations in this parameter will not significantly change the final values of the computed magnetic field strength. The resulting field strengths in the disk are consistent with typical values for a spiral galaxy
\citep{Beck_2001}. For the two blobs (NW and SE), $B_{\rm{tot}}$ was estimated by using a mean degree of polarization for the NW blob of
$p\cong40 \pm 10 \%$ derived from Fig.~\ref{degree_pol} and assuming a similar degree of polarization for the SE blob. The uncertainty in p introduces an error in $B_{\rm{tot}}$ of about $10\%$.

The intrinsic degree of polarization $p_0$ in the NW blob may be larger if Faraday depolarization occurs there,
possibly by $\Ha$ clouds similar to those observed in the north-east by \cite{Yoshida_2002}.
A turbulent field strength $B_\mathrm{turb} = \sqrt{B_\mathrm{tot}^2 - B_\mathrm{ord}^2} \cong 8\,\mu \rm{G}$,
a pathlength of about $2300\,\rm{pc}$ (Table~\ref{B_all_parts_table}), an rms thermal electron density of about $0.35\ccm$ in
$\Ha$ clouds of about $200\,\rm{pc}$ size and $2\times10^{-3}$ volume filling factor \citep{Yoshida_2002} lead to a
dispersion in rotation measure of $\sigma_\mathrm{RM} \cong 70 \radm$. The resulting depolarization at $6\,\rm{GHz}$
by internal Faraday dispersion \citep[e.g.][]{Sokoloff_1998,Arshakian_2011} is negligible ($p/p_0 \cong0.97$).
Faraday depolarization by hot gas in the halo of NGC\,4388 is even less significant because its electron density is only few times $10^{-3}\ccm$ \citep{Wezgowiec_Mach}.

The strengths of the ordered field obtained from the polarization map are lower limits, since this emission only represents the component of the magnetic vector in the plane of the sky. If the ordered field is inclined by, say, $30\degr$ with respect to the line of sight, its strength would increase by $11\%$.

\begin{figure}[t!]
\includegraphics[width=1\columnwidth]{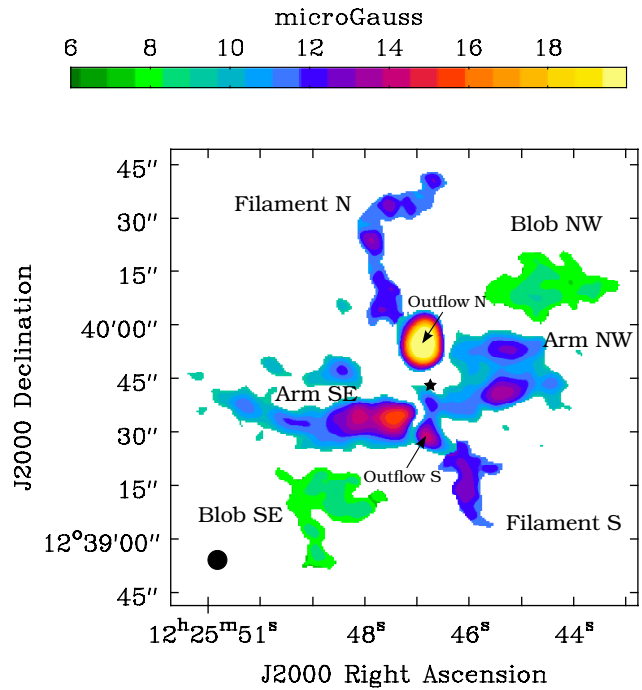}
\caption{\small{Strengths of the ordered field computed from the map of polarized intensities using the
revised equipartition formula by \citet{Beck_2005}. The nucleus of the galaxy is marked with a star.}}
\label{B_all_parts}
\end{figure}

Figure~\ref{B_all_parts} shows the strengths of the ordered magnetic field obtained from the map of polarized
intensities of NGC\,4388. The strength of the ordered magnetic field within the disk is comparable to that of other spiral galaxies. The highest ordered field strengths in the extraplanar region are observed in the vertical filaments. It is notable that the magnetic field in the NW and SE blobs is weaker than in the filaments
although the polarized intensities are similar. This is due to the different pathlengths chosen for each region (see Sect.~\ref{M82}).

%%%%%%%%%%%%%%%%%%%%%%% Table of Magnetic field values %

\begin{table}
\centering  % used for centering table
\begin{minipage}{\columnwidth}
\caption{Equipartition field strengths in NGC\,4388}              % title of Table
\begin{tabular}{l c c c c}  % centered columns (4 columns)
\hline\hline                        % inserts double horizontal lines
 &L (kpc)&$\alpha$ & $\rm{B_{ord}}\,(\mu \rm{G})$ & $\rm{B_{tot}}\,(\mu \rm{G})$ \\    % table heading
\hline                                   % inserts single horizontal line
Nucleus & 1.0 & 0.8 & - & 67 \\% inserting body of the table
Outflow N & 1.0 & 0.8 & 23 & 45 \\
Outflow S & 1.0 & 0.8 & 14 & 30 \\
Arm/disk SE & 1,~ 4 \footnote{The path-length through the spiral arm is used for polarization,\newline
\hspace*{1.8em} the path-length through the inclined disk for total intensity.  \label{first_footnote}} & 0.8 & 16 & 23 \\
Arm/disk NW & 1,~ 4 \footref{first_footnote}  & 0.8 & 13 & 21 \\
Blob NW & 2.3 & 1.0 & 9 & 11 \footnote{Estimated from the degree of polarization of Fig. \ref{degree_pol}.  \label{second_footnote}} \\
Blob SE & 2.3 & 1.0 & 9 & 11 \footref{second_footnote} \\
Filament N & 0.5 & 1.0 & 13 & - \\
Filament S & 0.5 & 1.0 & 13 & - \\
\hline                                            %inserts single line
\end{tabular}
\label{B_all_parts_table}
\end{minipage}
\end{table}

\section{Discussion}
\label{Interpretation}
\subsection{Nuclear outflow}
\label{Outflow}

We clearly detect the northern outflow extending from the center of NGC\,4388 in total intensity (Fig.~\ref{TotalI_PA}), while the southern counterpart is weaker.
Due to the inclination of the galaxy, the emission from the northern outflow travels through the disk on the way to the observer, contrary to the emission of the southern counterpart. It is common in AGN observations to see one side of the jet brighter than the other due to Doppler beaming \citep{Pearson,Kellerman}. The blue-shifted velocities of the $\Ha$ outflow in the north indicate that the northern outflow is pointing toward the observer. In those cases, the jet has an inclination toward the observer and it travels at relativistic speed. This high speed would imply a difference in size between northern and southern polarization filaments due to relativistic effects. However, the AGN jet cannot remain relativistic until kilo-parsec scales. A possible explanation for this configuration would imply an internal asymmetry in the ISM density close to the AGN core. This situation is supported by $\OIII$ observations \citep{Falcke} where there is a region of diffuse emission toward the south of the nucleus, possibly due to an interaction between the AGN outflow and the ISM gas.

In the northeastern part of the halo we can identify total power emission
associated with the outflows in $\Ha$ and $\OIII$ \citep{Yoshida_2002,Yoshida_2004} and X-rays \citep{Iwasawa}.

The northern outflow shows strongly polarized emission at the top with degrees of
polarization of $12\,\%$ (Fig.~\ref{degree_pol}), while the central part of the same structure, i.e., the nuclear region of the galaxy, reveals no
detectable polarization ($<1\,\%$). As the outflow is oriented almost perpendicular to the galaxy disk,
Faraday depolarization could occur in the northern (nearby) part of the disk that is located between the outflow and the observer.
Strong depolarization ($p/p_0 < 0.1$) at 6\,GHz by external Faraday dispersion requires a dispersion in
rotation measure of $\sigma_\mathrm{RM} > 400 \radm$, e.g.  by a diffuse ionized medium with an average thermal electron
density of $> 0.03\ccm$, assuming a turbulent field strength in the disk of $B_\mathrm{turb} \cong 15\,\mu \rm{G}$
(Table~\ref{B_all_parts_table}), a pathlength through the disk of about 5000\,pc, 100\,pc turbulence
length and a volume filling factor of 0.5. Such electron densities are typical for the ISM of spiral
galaxies \citep[e.g.][]{Beck_2007}. Alternatively, depolarization could occur by internal Faraday dispersion
in the outflow itself. For a turbulent field strength of $B_\mathrm{turb} \cong 50\,\mu \rm{G}$ and a
pathlength of 1000\,pc, an average internal thermal electron density of $> 0.05\ccm$ is needed, which is
a reasonable value.

\begin{figure*}
\plottwo{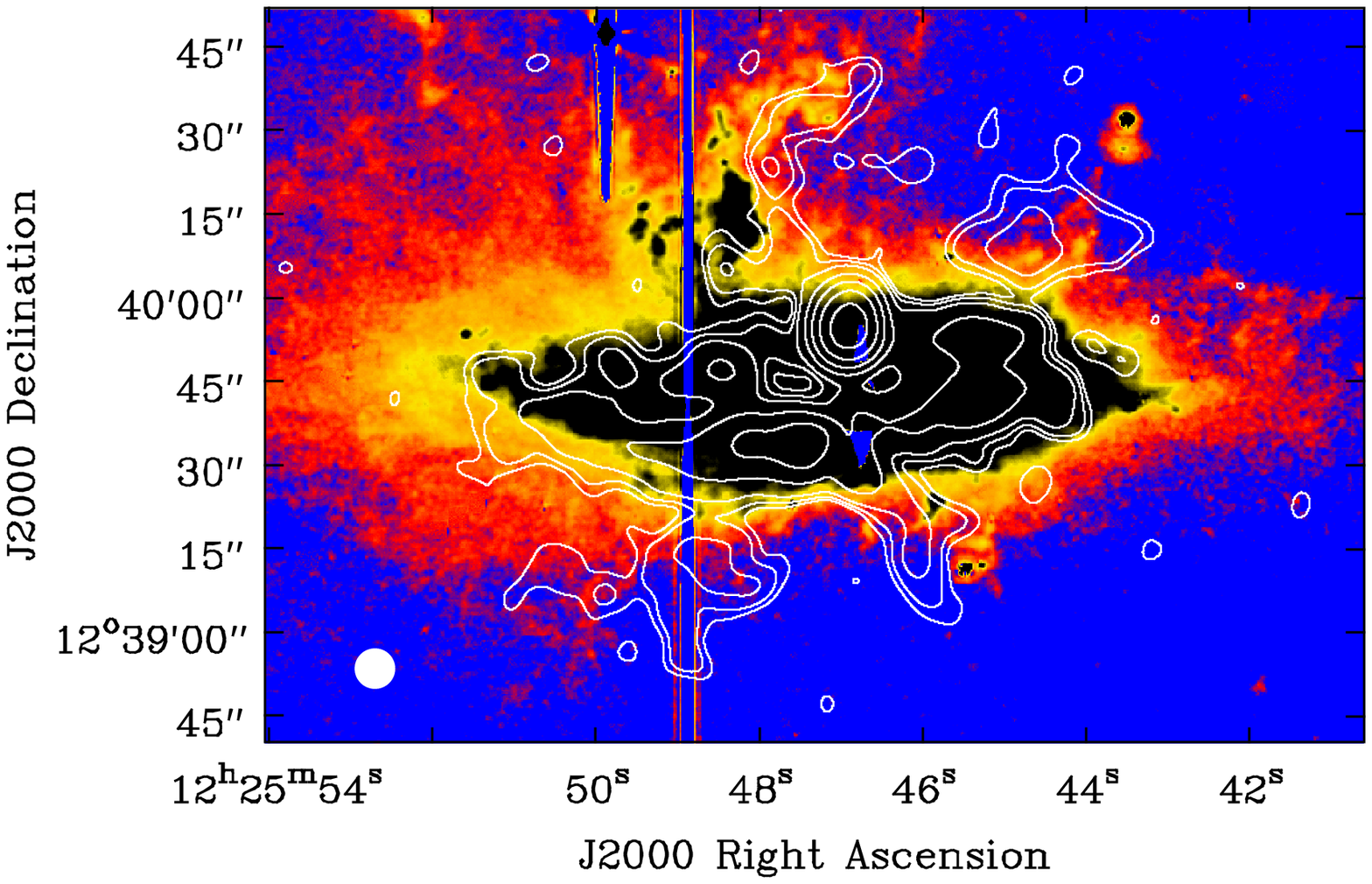}{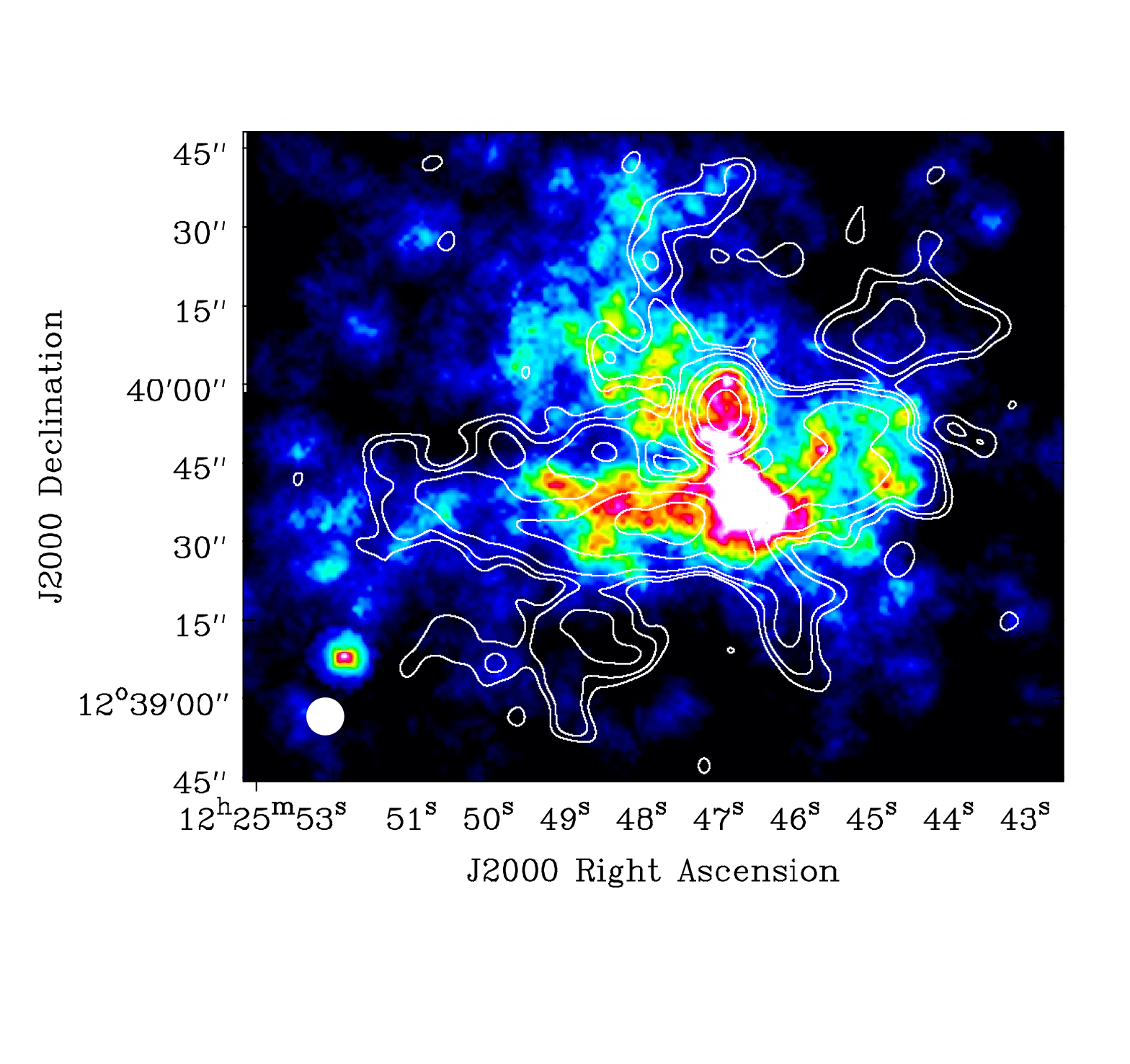}
\caption{\small{Left: Polarized emission obtained with RM Synthesis compared to $\Ha$ observations from
\citet{Yoshida_2002}. Contours of polarized emission are $(5, 6, 8, 16, 32, 64, 128)\,\times 2\,\mu\rm{Jy/beam}$.
The resolution is $7\arcsec\,\times\,7\arcsec$ and the rms noise is $2\,\mu\rm{Jy/beam}$. Right: Same polarized emission contours over CHANDRA X-ray map from \cite{Iwasawa}.}}
\label{Halpha_X-ray}
\end{figure*}

Thanks to the new polarized emission maps, we are able to identify for the first time the counter-outflow
toward the south. The main indication is the change of the orientation of the magnetic vectors from parallel
to the spiral arms of the disk to being oriented along the nuclear outflow in the southern part that is connected to the center of
the galaxy (Fig.~\ref{TotalI_PA}). This is a clear indication of a different structure in front
of the southern arm.
The degree of polarization in the southern outflow is $8\,\%$, similar to the outer part of the northern outflow.
Contrary to the central northern outflow, the southern outflow does show polarization,
probably because the southern spiral arm does not obstruct the view of the southern outflow by the observer.

\subsection{NGC4388: not an M82-like superwind}
\label{M82}

The particular configuration of filaments and blobs seen in polarized radio continuum and $\Ha$ emission of NGC\,4388 might suggest that these are different parts of the same event. Both filaments and blobs seem to form an hourglass-shaped structure. This kind of structure is typically seen in starburst galaxies. 
The prototypical starburst galaxy M\,82 shows such an hourglass shape in $\Ha$ emission. In such a case, strong star formation in the galactic center and subsequent supernova explosions eject gas into the galactic halo within a biconical structure \citep{Strickland_2002}. Hot tenuous gas expands as a superbubble into the halo pushing the extraplanar disk to larger galactic radii and compressing it. The compressed gas cools radiatively and becomes visible in many frequency ranges, leading to the observed hourglass shape. Due to gas compression the magnetic field is enhanced and aligned, giving rise to a large scale ordered magnetic field observed as polarized radio continuum emission (e.g. NGC\,5775 \citealt{Soida_2011}).
Within this picture the compressed shells should also be visible in X-rays (M\,82 \citealt{Ranalli_2008}, NGC\,253 \citealt{Dahlem_1997}, NGC\,5775 \citealt{Li_2008}).
Observations of the diffuse X-ray emission of M\,82 \citep{Strickland_1997} show that the superbubble in this galaxy is partly filled with hot ($\gtrsim 10^6\,\rm{K}$) gas. The radio continuum emission of M\,82 \citep{Adebahr_2013} is co-spatial with the $\Ha$ and the X-ray emission. In NGC\,5775 the diffuse X-ray emission \citep{Li_2008} fills the radio continuum superbubble on the southwestern quadrant of the disk \citep{Soida_2011}.
The observed correlation between X-rays, $\Ha$, and radio continuum indicates that the hot X-ray plasma is closely linked to cooler and denser gas detected in $\Ha$ and to CREs. Such filaments extend into the galactic halo up to $\cong 10\,\rm{kpc}$ in the case of NGC\,253 \citep{Strickland_2002}. 

In NGC\,4388 the situation is different. The X-ray emission (see right panel of Fig. \ref{Halpha_X-ray}) is not symmetric as expected from superwind models. Compared to the X-ray emission of NGC\,253 we do not see enhanced extraplanar X-ray emission south of the galactic disk, in the area where the base of the bubble should be found. North of the galactic disk there is extraplanar X-ray emission. However, it is very asymmetric with almost all X-ray emission coming from the northeastern side of the galactic disk. In addition, the polarized radio continuum emission shows a filamentary structure as expected by a galactic outflow \citep{Soida_2011} only in the northeastern and the southwestern quadrants. Moreover, the observed SE and NW blobs are elongated parallel to the disk. Therefore, we conclude that the observed spatial distribution of polarized radio continuum, $\Ha$ and X-ray emission is not consistent with a symmetric superwind scenario. In Sect.~\ref{Blobs} we show that NGC\,4388 hosts a less prominent galactic wind probably originating from the spiral arms.
\newline
\newline
\subsection{Origin of the polarization filaments}
\label{Arcs}

We detect faint polarized emission filamentary in structure extending beyond the end of both sides of the outflow. There are
striking similarities in morphology between these features and the $\Ha$ outflow (left panel of Fig.~\ref{Halpha_X-ray}). The polarized radio continuum filaments are offset from the $\Ha$ plumes toward the minor axis of the galaxy.
The projected lengths of these filamentary structures are about 50\arcsec\ $(4.1\,\rm{kpc})$ and 20\arcsec\ $(1.7\,\rm{kpc})$
for the northern and southern parts, respectively. In both cases the deconvolved half power width is about 3.9\arcsec\
(0.3\,kpc). The symmetry of the two filaments with respect to the galaxy center suggests a collimated outflow perhaps driven by a radio jet. The position angle of the filament is about $20^{\circ}-30^{\circ}$, i.e. it is not perpendicular to the disk plane. The magnetic field vectors
show an ordered field along both filaments (see Fig.~\ref{PI_PA}). None of these external features are seen in total intensity due to the
lower noise in the polarization maps.

The magnetic field vectors in the southern part connect the nucleus with the southern polarization filament through an apparently wiggling structure. This change of direction could be due to precession event of the AGN jet. In that case, the polarization filaments we see at kiloparsec scales are the continuation of a structure that originates in the nucleus. The fact that there is a drastic drop in polarized intensity between the nuclear outflows and the filaments could be the result of an abrupt change in the direction of the outflow giving rise to this change in polarized intensity. In that case, what we see are different parts of the same event. The origin of the filaments of NGC\,4388 will be discussed in more detail in a future paper.

\subsection{Origin of the polarization blobs in the halo}
\label{Blobs}

We observe two blobs of polarized emission far out in the galactic halo (about $2\,\rm{kpc}$ from the disk plane), in the northwestern and southeastern regions above and below the disk, with magnetic field vectors forming an almost vertical field structure (Fig.~\ref{TotalI_PA} and \ref{PI_PA}). The magnetic field vectors within the disk and in the halo are perpendicular to each other, so that there are two areas between the blobs and the disk of
low polarization due to geometrical depolarization within the beam.
The vertical field in the blobs can be part of a large-scale halo field or field loops stretched by a
strong galactic wind (see Sect.~\ref{Ram pressure}). These polarized blobs remain undetected at other frequencies, e.g., optical, $\Ha$, or X-rays. However, some of the channel maps of the $\HI$ cubes from \cite{Chung} show structures that extend from the outskirts of the western part of the $\HI$ disk in the northeastern direction, coincident with the NW blob, indicating a possible relation to the blobs detected in polarization.
We interpret this as a galactic outflow from the spiral arms. The fact that the polarized blobs are detected close to both spiral arms in projection ($\approx 15\,\arcsec$) may indicate that they are related to the disk. Furthermore, there is X-ray (see right panel of Fig.~\ref{Halpha_X-ray}), polarized and total power emission extending $\approx 10\,\arcsec$ south of the southeastern spiral arm and north of the northwestern spiral arm. If so, possibly the CREs are transported by the galactic wind from the spiral arm. In this case, we estimate the speed of the galactic wind using CREs synchrotron lifetime \citep[e.g.][]{Beck_2013}:

\begin{equation}
t_{\rm{syn}} \cong 1.06 \times 10^{9} \,\rm{yr}\left(\frac{\textit{B}}{\mu\rm{G}}\right)^{-\frac{3}{2}} \left(\frac{\nu}{\rm{GHz}}\right)^{-\frac{1}{2}}\,\cong (12\pm3)\,\rm{Myr}
\end{equation}
where $B \cong 11\,\mu\rm{G}$ is the total magnetic field (Table~\ref{B_all_parts_table}) with an estimated error of about $2\,\mu\rm{G}$, and
$\nu \cong 6.0\,\rm{GHz}$ is the central observation frequency.
The distance between the spiral arm and the outer edge of the blobs (at the level of $5\sigma$)
is about $35\arcsec \cong 2.9\,\rm{kpc}$ for the NW blob and about $45\arcsec \cong 3.7\,\rm{kpc}$ for the SE blob.
To reach this height, CREs have to travel with an average velocity of $v \cong (270\pm70)\,\rm{km\,s^{-1}}$,
which is typical for the speed of galactic winds \citep[e.g.][]{Heesen,Arribas}.
The true speed in the blob depends on the velocity profile of the outflow and could be several times larger,
at least by a factor of two for constant acceleration.

The velocity of about $270\,\rm{km\,s^{-1}}$ refers to the component of the galactic wind in the sky plane.
With a disk inclination of $79\degr$ and assuming an outflow velocity vertical to the disk plane, we can estimate the component parallel to the line of sight of that vector and
compare it to the observed radial velocities in $\Ha$. We estimate the radial velocity of the outflow to be $\approx\,60\,\rm{km\,s^{-1}}$, which is close to the value observed in $\Ha$ \citep{Veilleux}.
Although there is no stringent reason for the same velocity of the hot (or CREs) and the $\Ha$-emitting cool gas flows, it is remarkable that these two components have similar velocities.

The magnetic field computed by the equipartition
formula ($11\,\mu\rm{G}$) could be underestimated. The energy spectrum of CREs propagating into
the halo is steepened by energy losses and hence is not proportional to the proton spectrum, i.e. the
proton-to-electron ratio K is significantly larger than 100 as assumed in Sect.~\ref{equi}
(see discussion in \citealt{Beck_2005}).
In this case we would get a shorter synchrotron lifetime and therefore a larger velocity of the outflow.
A better estimate of the outflow speed needs knowledge of the synchrotron spectral index as a function of
height above the plane, which cannot be obtained with the present data.

It is surprising that there is no counterpart of both blobs on the opposite sides of the disk in any other wavelength. The two blobs might be due to an earlier AGN-related outflow, because AGNs can turn off and on as well as change direction. However, the age of a previous outflow would be much beyond the synchrotron lifetime (see above). In the case of a continuous galactic wind expanding from the disk, one would expect that the outflow of particles and magnetic field is equally strong on both sides of the spiral arm.
The asymmetric distribution suggests that the outflow is not homogeneous, but emerges from individual star-forming
complexes that are located not exactly in the disk plane. 
The non-detection of the NE counterpart on the other hand, could be caused by intrinsic depolarization due to the overlapping with the material coming from the nuclear outflow in that area.

We conclude that NGC\,4388 hosts a collimated nuclear outflow and a galactic wind originating mainly from the spiral arms.

\subsection{Halo pressure}
\label{Halo_pressure}

The halo pressure can have several components: thermal gas pressure, magnetic pressure, cosmic ray pressure, and ram pressure from the galactic outflow. The thermal pressure of the hot gas of $P_{\rm{th}}= nkT\,\cong (3 \pm 2) \times 10 ^ {-12}\,\rm{dyn\,cm^{-2}}$ is estimated using a temperature of $kT=(0.5\pm 0.3)\,\rm{keV}$ and a density of $n\cong(4\pm 2)\times10^{-3}\,\rm{cm^{-3}}$ from \cite{Wezgowiec_Mach}, Table 3, interpolated between disk and tail, assuming a volume filling factor of 0.5. With an expansion velocity of $(270\pm70)\,\rm{km\,s^{-1}}$, the ram pressure of the galactic wind is $P_{\rm{wrp}}= 0.61\,n\,\rm{m_H}\,v^2\cong(3 \pm 2)\times10^{-12}\,\rm{dyn\,cm^{-2}}$. With an equipartition  magnetic field strength of $B \cong (11\pm2)\,\mu \rm{G}$ the magnetic pressure\footnote{The magnetic vectors in the blobs show a magnetic field which is oriented along the wind and therefore the magnetic pressure term is smaller in this direction.} is $P_{B}\cong B^2/(8\pi)\cong(5\pm2)\times 10^{-12}\,\rm{dyn\,cm^{-2}}$. Thus, there is approximate equipartition between the different pressure components (except for the cosmic ray pressure that is one third of the magnetic pressure in case of energy equipartition). If the pressure components can be simply added, the total pressure is about $1.3\times 10^{-11}\,\rm{dyn\,cm^{-2}}$. If, however, the pressure components are not co-spatial or some of the pressure fluctuations are anticorrelated,  they should not be added to a total pressure value.
Hence, the pressure in the halo is in the range of $P_{\rm{halo}}\cong3-13 \times 10^{-12}\,\rm{dyn\,cm^{-2}}$.

The equipartition between the ram pressure of the outflow
and the thermal pressure can be understood by the interplay between
shocks and the ambient halo gas, which keeps the outflow
velocity close to the sound speed of the hot halo gas.

A large-scale magnetic field may exist in the entire halo of NGC\,4388 (a magnetosphere), but is observable only in regions into
which a sufficiently large number of CREs is supplied from the disk by an outflow. Evidence for a large-scale regular field in the halo has to come from measurements of Faraday depth at lower frequencies with good resolution in Faraday space, i.e. a wide coverage in $\lambda^2$ space, which is not the case for the present observations centered
at $6\,\rm{GHz}$. Polarization observations in L band (centered at 1.5\,GHz) are strongly affected by Faraday depolarization,
hence VLA S band (centered at 3\,GHz) seems more promising. A lack of Faraday rotation around $3\,\rm{GHz}$ would support the idea of
stretched field loops.

\subsection{ICM ram pressure}
\label{Ram pressure}

In the first part of this subsection the ICM ram pressure is estimated via the H{\sc i} stripping radius, the thermal pressure of the resisting ISM and via a detailed comparison between observations and dynamical simulations. Magnetic pressure may help to resist ICM ram pressure. Its importance is
investigated in the second part of this subsection. As a consistency check, the lower limit of the ICM ram pressure can be estimated via ionized
high-velocity filaments that are accelerated by ram pressure. All ram pressure estimates yield values in excess of the pressure in the galactic halo.
We suggest that the observed ICM clumpiness can account for this discrepancy, i.e. the galaxy is moving a portion of the clumpy ICM where the
local density is several times lower than value predicted by a continuous smooth ICM distribution.

\subsubsection{Ram pressure estimates}
\label{Ram pressure estimates}

NGC\,4388 is moving to the southwest. Therefore, we expect ram pressure compression in the southern part of the galactic disk. \citet{Vollmer_2009} described a scenario where the peak of ram pressure
for this galaxy has already passed but in which the galaxy is still being affected in a considerable way by the ICM.
The peak of the ram pressure in NGC\,4388 occurred $190\pm30\,\rm{Myr}$ ago \citep{Pappalardo} having $P_{\rm{max}} \approx  5 \times 10^{-11}\,\rm{dyn}\,\rm{cm}^{-2}$ \citep{Vollmer_2003}. At this time step the absolute velocity with respect to the
cluster core is estimated to be $v \approx 2000\,\rm{km\,s^{-1}}$ and the maximum ICM density $n_{\rm{ICM}} \approx 1 \times 10^{-3}\, \rm{cm}^{-3}$ (\citealt{Vollmer_2003} erroneously gave a maximum pressure three times higher than the actual value). The model predicts that the galaxy is now
affected by a ram pressure of $1/5$ of the peak ram pressure occurred in the past. This
means that at the present day this value would be
$P \approx 1 \times 10^{-11}\,\rm{dyn}\,\rm{cm}^{-2}$.
\cite{Roediger_2006} performed high-resolution 3D hydrodynamical simulations, using different parameters of densities and velocities, to reproduce the stripped gas as result of the interaction between a massive galaxy and the ICM. In this study, the authors were also able to reproduce outflows of about 100\,kpc extension. From all the scenarios considered, they concluded that the most similar one to the case of NGC\,4388 can be described by a ram pressure of a few
$10^{-11}\,\rm{dyn}\,\rm{cm}^{-2}$. However, improved hydrodynamical models by \cite{Roediger_2008} could not
explain the shape of the gas tails of NGC\,4388, possibly because the ICM is inhomogeneous or the outflow
influences the interaction between the galaxy and the ICM. All these ram pressure estimates are based on the assumption that the spatial distribution of the ICM is continuous and smooth.

As a consistency check we apply the method of \cite{Vikhlinin_2001} to calculate the ICM density. The ratio between the thermal pressure at the stagnation point $P_0$ and the thermal pressure of the freely streaming ICM $P_{\rm{ICM}}$ is a function of the galaxy's Mach number $\mathcal{M}$:
\begin{equation}
\frac{P_0}{P_{\rm{ICM}}}= \left(\frac{\gamma + 1}{2}\right)^{(\gamma + 1)/(\gamma - 1)}\,\mathcal{M}^2\left[\gamma - \frac{\gamma - 1}{2\mathcal{M}^2}\right]^{-1/(\gamma - 1)},
\label{Vikhilin}
\end{equation}
where $\gamma = 5/3$ is the adiabatic index of the monoatomic gas. With an ICM temperature of $2.2\,\rm{keV}$ \citep{Bohringer_1994} the ICM sound speed is approximately $c_{\rm{ICM}}\cong750\,\rm{km\,s^{-1}}$. We assume a galaxy velocity within the Virgo cluster of $v_{\rm{gal}}\cong1700\,\rm{km\,s^{-1}}$. The Mach number thus is $\mathcal{M}\cong2.3$, leading to a pressure ratio of $P_0/P_{\rm{ICM}} \cong 8$. Assuming equilibrium between the thermal pressure of the hot gas halo of the galaxy and the ICM gas at the stagnation point, we can replace $P_0$ by the halo pressure. With the thermal pressure of the halo $P_{\rm{th}}$ from Sect.~\ref{Halo_pressure}, we obtain an ICM thermal pressure of $P_{\rm{ICM}}\cong4\times 10^{-13}\,\rm{dyn}\,\rm{cm}^2$ and an ICM density of $n_{\rm{ICM}}\cong1\times 10^{-4}\,\rm{cm}^{-3}$. This agrees with the density derived from X-rays \citep{Urban} and from the Sunyaev-Zel'dovich (SZ) signal measured by \textit{Planck} \citep{Planck_2015}, while the dynamical model yields an ICM density of $n_{\rm{ICM}}^{\rm{model}}\cong3\times 10^{-4}\,\rm{cm}^{-3}$.

The model density thus seems to be overestimated by a factor of about $3$. However, magnetic fields are not taken into account in the ICM pressure estimate of Eq.~\ref{Vikhilin}. Including magnetic fields will increase the estimated ICM ram pressure and thus the estimated ICM density. In Sect.~\ref{ICM clumpiness} we show that ICM clumping might also mitigate the discrepancy between the two ICM pressure estimates.

Radio continuum observations of other Virgo galaxies affected by ram pressure like NGC\,4402 or NGC\,4501 also show signs of interaction with the ICM: both galaxies show a truncated gas disk \citep{Chung} with enhancements of polarization and sharp edges of the total power emission on one side of the galactic disk \citep{Vollmer_2007,Vollmer_2010}. \cite{Vollmer_2008} estimated a ram pressure of $P \approx 8.2 \times 10^{-12}\,\rm{dyn}\,\rm{cm}^{-2}$ for NGC\,4501. For NGC\,4402 we can estimate an upper limit of ram pressure by assuming that the western $\HI$ tail is pushed to higher galactic latitudes. According to \cite{Gunn_1972} the gravitational  restoring force balances ram pressure: $P_{\rm{ram}}= \Sigma v_{\rm{rot}}^2 / \textit{R}$. With a rotation velocity of $v_{\rm{rot}}\cong150\,\rm{km\,s^{-1}}$, a gas surface density of $\Sigma\cong10\,\rm{M_{\odot}}\,\rm{pc^{-2}}$, and an $\HI$ radius of $R_{\HI}\cong7.7\,\rm{kpc}$ the ram pressure of NGC\,4402 is $P_{\rm{ram}}\cong 3.2\times 10^{-11}\,\rm{dyn\,cm^{-2}}$. If the $\HI$ tail is ram pressure stripped gas that now falls back to the galactic disk the actual ram pressure would be a factor of few lower than our estimate.

The estimated ram pressure of NGC\,4388 is somewhat smaller than that of NGC\,4402 and similar to that of NGC\,4501. Since these galaxies show sharp edges of the radio continuum distribution and polarized ridges on one side of the galactic disk, we would expect to see the same phenomenon in NGC\,4388.

\subsubsection{Magnetic pressure}
\label{Magnetic pressure}

In contrast to NGC\,4402 or NGC\,4501, the radio halo of NGC\,4388 does not show a sharp southern edge. In addition, the polarized southwestern blob clearly belongs to the radio halo and thus implies that it is not compressed by ram pressure. The southern filament with a stronger magnetic field seems also unaffected by ram pressure. As observed in NGC\,4569 \citep{Chyzy_2006} the pressure of a galactic wind superbubble of NGC\,4388 (Sect.~\ref{Halo_pressure}) may resist ram pressure.

The magnetic pressure of a strong magnetic field may prevent ram pressure from removing the magnetic features seen in the halo. For that
reason we estimate the magnetic field needed to balance the model ram pressure of $P_{\rm{ram}} \approx 1 \times 10^{-11}\,\rm{dyn}\,\rm{cm}^{-2}$.
Setting this ram pressure equal to the magnetic energy density,
a total magnetic field of about $16 \, \mu\rm{G}$ is obtained. The value obtained from our observations
in the southern blob is $(11\pm2) \, \mu\rm{G}$ (Table~\ref{B_all_parts_table}), smaller than what is needed to
balance ram pressure, so that we would not expect any extensions of the magnetic field toward the southern halo.

We also consider the possibility that energy equipartition between total cosmic rays and total
magnetic fields is not valid in the halo, e.g. because the magnetic field in the expanding wind flow is
highly ordered. A close coupling of the cosmic rays to the field needs scattering at field irregularities,
which may be less efficient in the halo. As a result, the cosmic rays may stream with respect to the
field with a velocity higher than the Alfv\'en speed.

A field strength resisting the ram pressure of the model by \citet{Vollmer_2009} would mean that the magnetic
energy density has to be larger than the equipartition value by at least a factor of $(16\,\mu\rm{G}/11\,\mu\rm{G})^2\cong2$. To provide the same
synchrotron intensity as in the equipartition case, the energy density of the total cosmic rays has to be
lower by a factor of $2^{-(1+\alpha)}\approx0.25$ (for a synchrotron spectral index $\alpha \cong 1$).
The corresponding ratio between the magnetic and cosmic-ray energy densities is $\gtrapprox2^{(2+\alpha)}\approx8$,
which is an extreme deviation from equipartition, causing dynamical effects that lead to recovery of equilibrium.

Another argument against a super-equipartition field of $16\,\mu\rm{G}$ field in the halo comes
from the synchrotron lifetime of CREs that would decrease to about $7\,\rm{Myr}$. An unrealistically high average
outflow speed of $500\,\rm{km\,s^{-1}}$ would be needed to reach the height of the blobs above the disk plane.

We conclude that the estimates of the field strengths in Table~\ref{B_all_parts_table} are realistic. Either the ram pressure cannot be balanced by the magnetic pressure alone or the modeled ram pressure is overestimated.

\subsubsection{Ionized high velocity filaments}
\label{Ionized high velocity filaments}

In the following, we will estimate a lower limit for the ram pressure needed to accelerate clumps of ionized gas observed by \cite{Yoshida_2004}. The ``West High Velocity Filaments'' contain several subclumps which are located east of the galactic disk outside the faint $\Ha$ halo. The radial velocities of the clumps in the filaments are between $-250$ and $-360\,\rm{km\,s^{-1}}$ with respect to the systemic velocity of NGC\,4388. The radial components of the rotation velocity in these regions have positive values. Since the galaxy is leaving the cluster, the ionized clumps are decelerated by ram pressure of the ICM to negative velocities. The acceleration of a gas clump is given by $a= P_{\rm{ram}}/\Sigma$, where $\Sigma$ is the column density of the clump. The acceleration is approximately $a = \Delta \textit{v} / \Delta \textit{t} = (\Delta \textit{v})^2 /\textit{L}$, where $L$ is the distance over which the clump is accelerated. We assume $\Delta \textit{v} \geq 350\,\rm{km\,s^{-1}}$ and $L \cong 5\,\rm{kpc}$. For the column density we take the mean gas density\footnote{This implies a volume filling factor of $10^{-3}$} of $n_{\rm{clump}}\cong0.04\,\rm{cm}^{-3}$ and a size of $s = 100\,\rm{pc}$ given by \cite{Yoshida_2002}. In this way we estimate the minimum ram pressure to accelerate the ionized clouds to the observed velocities to be $P_{\rm{low}}\geq\rho\, s (\Delta \textit{v})^{2} / \textit{L} \cong 1.6\times 10^{-12}\,\rm{dyn\,cm^{-2}}$, a factor of $2$ lower than the individual halo pressure terms estimated in Sect.~\ref{Halo_pressure}. This lower limit is consistent with ram pressure being equal or somewhat lower to the halo pressure due to the galactic wind.

\subsubsection{ICM clumpiness}
\label{ICM clumpiness}

From the previous sections, there are indications that the ICM ram pressure from the numerical models (\citealt{Vollmer_2009}, \citealt{Roediger_2006}) may be overestimated by a factor of several at this cluster radius.

X-ray spectroscopy of the Virgo cluster \citep{Urban} indicates a drastic drop of temperature and metallicity beyond a radius of $r\,\cong 450\,\rm{kpc}$ from the cluster
center. A natural explanation for this decrease is clumping of the ICM that sets in at that radius. Based on the detailed comparison between observations and dynamical models \cite{Vollmer_2009} determined the 3D distance of the Virgo spiral galaxies. With respect to the cluster center, the 3D distance of NGC4388 is estimated to be $420\,\rm{kpc}$. This distance compares well with the distance where the ICM becomes clumpy. We note that due to clumpiness the actual ICM ram pressure acting on NGC\,4388 might be up to a factor of two lower than what is expected from a model including a smooth continuous ICM.

The \textit{Planck} measurements of the Virgo cluster 
SZ effect \citep{Planck_2015} show that radially averaged ICM clumping cannot be strong at the cluster distance of $420\,\rm{kpc}$. Based on numerical simulations, typical radially averaged clumping factors $C=<n_{e}^{\,\,2}>/<n_e>^2$ of unrelaxed clusters are $C=1.5-2$ \citep{Zhuravleva_2013}. Locally, the density can be enhanced by a factor of $3-10$. However, locations with densities exceeding three times the mean density are extremely rare (see Fig.~3 of \citealt{Zhuravleva_2013}).

With a galaxy velocity of $1700\,\rm{km\,s^{-1}}$ and an ICM density of $n_{\rm{ICM}}\cong1\times 10^{-4}\,\rm{cm}^{-3}$ \citep{Urban}, the ICM ram pressure is $P_{\rm{ram}}\cong 3\times 10^{-12}\,\rm{dyn\,cm^{-2}}$ which is consistent with the thermal halo pressure (Sect.~\ref{Halo_pressure}). If the magnetic and CR pressures are co-spatial with the thermal halo pressure, the actual ICM ram pressure can exceed $3\times 10^{-12}\,\rm{dyn\,cm^{-2}}$. On the other hand, the dynamical model predicts an ICM ram pressure of $1\times 10^{-11}\,\rm{dyn\,cm^{-2}}$. Thus, ICM clumping cannot be excluded and the actual ICM ram pressure acting on NGC\,4388 might well be a few times higher than the radially averaged ICM density estimated from X-ray and SZ observations.

Current dynamical simulations of ram pressure stripping events assume a continuous ICM distribution (\citealt{Vollmer_2001}, \citealt{Roediger_2006}, \citealt{Tonnesen_2009}). The influence of a clumpy ICM on these simulation depends on the stage of the interaction: a clumpy ICM distribution significantly affects only the beginning and the end of the simulations when the distance to the Virgo cluster center of the galaxy exceeds $\approx 500\,\rm{kpc}$. If the galaxy is observed close to or after peak ram pressure the changes of the simulation results with respect to a continuous ICM distribution are expected to be minor. However, if the galaxy is observed at the beginning of the
ram pressure stripping event or more than $200\,\rm{Myr}$ after peak ram pressure, the clumpy ICM distribution will have a significant influence on gas distribution and velocity.

\section{Summary}
\label{summary}
Our new VLA $6\,\rm{GHz}$ broad-band observations of the edge-on Virgo cluster galaxy NGC\,4388 allowed us to reach unprecedented low noise levels, revealing striking new details of this object. The polarized emission obtained for the first time for this galaxy with RM synthesis, show extensions of the magnetic field toward the outskirts of the galaxy, indicating a connection between disk and halo. Two polarized filamentary structures appear at the end of both northern and southern nuclear outflows which correlate with $\Ha$ and X-ray observations. The change in orientation of the magnetic vectors in the southern spiral arm reveals for the first time the southern nuclear outflow. Furthermore, two horizontally extended blobs of polarized emission are observed in the halo, about $2\,\rm{kpc}$ above and below the northeastern and southwestern spiral arms, respectively. Within these blobs, the ordered magnetic field is oriented perpendicular to the galactic disk.

The comparison between multi-wavelength observations of prototype galactic winds shows that NGC\,4388 does not host a symmetric central galactic wind.
However, we suggest that, together with $\Ha$, X-ray, and total power emission being located closer to the spiral arms within the halo, the blobs of polarized emission trace a galactic wind, which most likely originate from separate sources in the spiral arms. In such a scenario the CREs travel from the spiral arms into the halo, reaching distances of up to $\cong 3.3\,\rm{kpc}$ in the plane of the sky. Assuming equipartition between CR particles and the magnetic field we estimate a total magnetic field strength for the different parts of the galaxy taking into account individual spectral indices and pathlengths. In particular, our estimate of the total magnetic field strength for the polarization blobs is $(11\pm2)\,\mu\rm{G}$. With this magnetic field strength, the synchrotron lifetime of electrons in the polarized blob is $\cong (12\pm3)\,\rm{Myr}\,$. For traveling a distance of $\cong 3.3\,\rm{kpc}$, the average outflow velocity of those particles is $(270\pm70)\,\rm{km\,s^{-1}}$, which agrees with the typical speed of a galactic wind expanding from the spiral arms into the halo.

Another edge-on Virgo galaxy, NGC\,4402, shows a sharp edge in the radio continuum emission which is interpreted as a compressed galactic halo. The observed symmetry of the polarized halo features in NGC\,4388 excludes a compression of the halo gas by ICM ram pressure. We estimate the halo magnetic pressure and the ram pressure of the galactic wind to be $P_{\rm{halo}}\approx 3 \times 10^{-12}\,\rm{dyn\,cm^{-2}}$. This pressure is comparable to the thermal pressure derived from X-ray observations.

The estimate of the ICM ram pressure based on the galaxy velocity from dynamical models and a radially averaged ICM density profile from X-ray observations yields $P_{\rm{ICM}}\cong3\times 10^{-12}\,\rm{dyn}\,\rm{cm}^{-2}$, in agreement with our estimate of the thermal halo pressure. If magnetic fields are ubiquitous in the halo, the halo pressure estimate increases.

Allowing for ICM clumping, the actual ram pressure acting on NGC\,4388 might be up to few times higher than what is expected based on a given galaxy velocity and an ICM density determined from X-ray and SZ observations.

NGC\,4388 besides NGC\,4569 \citep{Chyzy_2006} is the second galaxy in the Virgo cluster which shows a galactic outflow resisting ICM ram pressure. The detection of a radio halo around other cluster spiral galaxies could be used for an estimate of ICM density and ram pressure within a factor of a few.

Further observations at lower frequencies (GMRT, LOFAR, SKA) will reveal stronger total intensity emission from the halo of this galaxy allowing us to do a spectral index analysis and therefore better constrain the aging
of the CREs. Measurements of the large-scale structure of the halo magnetic field in NGC\,4388 with help of polarization data are needed, e.g. with the VLA S band where good resolution in Faraday space can be reached.
$\Ha$ observations are also crucial to constrain the velocities within the polarized blobs. We expect to see blue-shifted emission in the wind of the southern part and red-shifted in the northern part due to 
the inclination of the galaxy towards the observer.

\acknowledgments

This work has used the Karl G. Jansky Very Large Array operated by The National Radio Astronomy Observatory (NRAO). The NRAO is a facility of the National Science Foundation operated under cooperative agreement by Associated Universities, Inc.

We thank the anonymous referee for useful comments. We also thank Aritra Basu, Marcus Br\"uggen, Klaus Dolag, Christian Fendt, Stefanie Komossa, and Robert Laing for useful discussions.

We acknowledge support by project DFG RU1254.

\bibliographystyle{apj} % style aa.bst
%\bibliography{n4388.bib} % your references Yourfile.bib

\begin{thebibliography}{}
\expandafter\ifx\csname natexlab\endcsname\relax\def\natexlab#1{#1}\fi

\bibitem[{{Adebahr} {et~al.}(2013){Adebahr}, {Krause}, {Klein},
  {We{\.z}gowiec}, {Bomans}, \& {Dettmar}}]{Adebahr_2013}
{Adebahr}, B., {Krause}, M., {Klein}, U., {et~al.} 2013, \aap, 555, A23

\bibitem[{{Arribas} {et~al.}(2014){Arribas}, {Colina}, {Bellocchi}, {Maiolino},
  \& {Villar-Mart{\'i}n}}]{Arribas}
{Arribas}, S., {Colina}, L., {Bellocchi}, E., {Maiolino}, R., \&
  {Villar-Mart{\'i}n}, M. 2014, \aap, 568, A14

\bibitem[{{Arshakian} \& {Beck}(2011)}]{Arshakian_2011}
{Arshakian}, T.~G., \& {Beck}, R. 2011, \mnras, 418, 2336

\bibitem[{{Beck}(2001)}]{Beck_2001}
{Beck}, R. 2001, \ssr, 99, 243

\bibitem[{{Beck}(2007)}]{Beck_2007}
---. 2007, \aap, 470, 539

\bibitem[{{Beck} \& {Krause}(2005)}]{Beck_2005}
{Beck}, R., \& {Krause}, M. 2005, Astronomische Nachrichten, 326, 414

\bibitem[{{Beck} \& {Wielebinski}(2013)}]{Beck_2013}
{Beck}, R., \& {Wielebinski}, R. 2013, in {Planets, Stars and Stellar
  Systems.~Volume 5: Galactic Structure and Stellar Populations}, ed. T.~D.
  {Oswalt} \& G.~{Gilmore}, 641

\bibitem[{{B{\"o}hringer} {et~al.}(1994){B{\"o}hringer}, {Briel}, {Schwarz},
  {Voges}, {Hartner}, \& {Tr{\"u}mper}}]{Bohringer_1994}
{B{\"o}hringer}, H., {Briel}, U.~G., {Schwarz}, R.~A., {et~al.} 1994, \nat,
  368, 828

\bibitem[{{Brentjens} \& {de Bruyn}(2005)}]{Brentjens}
{Brentjens}, M.~A., \& {de Bruyn}, A.~G. 2005, \aap, 441, 1217

\bibitem[{{Cayatte} {et~al.}(1990){Cayatte}, {van Gorkom}, {Balkowski}, \&
  {Kotanyi}}]{Cayatte}
{Cayatte}, V., {van Gorkom}, J.~H., {Balkowski}, C., \& {Kotanyi}, C. 1990,
  \aj, 100, 604

\bibitem[{{Chung} {et~al.}(2009){Chung}, {van Gorkom}, {Kenney}, {Crowl}, \&
  {Vollmer}}]{Chung}
{Chung}, A., {van Gorkom}, J.~H., {Kenney}, J.~D.~P., {Crowl}, H., \&
  {Vollmer}, B. 2009, \aj, 138, 1741

\bibitem[{{Chy{\.z}y} {et~al.}(2006){Chy{\.z}y}, {Soida}, {Bomans}, {Vollmer},
  {Balkowski}, {Beck}, \& {Urbanik}}]{Chyzy_2006}
{Chy{\.z}y}, K.~T., {Soida}, M., {Bomans}, D.~J., {et~al.} 2006, \aap, 447, 465

\bibitem[{{Dahlem}(1997)}]{Dahlem_1997}
{Dahlem}, M. 1997, \pasp, 109, 1298

\bibitem[{{Falcke} {et~al.}(1998){Falcke}, {Wilson}, \& {Simpson}}]{Falcke}
{Falcke}, H., {Wilson}, A.~S., \& {Simpson}, C. 1998, \apj, 502, 199

\bibitem[{{Giroletti} \& {Panessa}(2009)}]{Giroletti}
{Giroletti}, M., \& {Panessa}, F. 2009, \apjl, 706, L260

\bibitem[{{Gunn} \& {Gott}(1972)}]{Gunn_1972}
{Gunn}, J.~E., \& {Gott}, I. J.~R. 1972, \apj, 176, 1

\bibitem[{{Heald}(2009)}]{Heald_2009}
{Heald}, G. 2009, in {IAU Symposium}, Vol. 259, {IAU Symposium}, ed. K.~G.
  {Strassmeier}, A.~G. {Kosovichev}, \& J.~E. {Beckman}, 591--602

\bibitem[{{Heald}(2015)}]{Heald_2015}
{Heald}, G. 2015, in {Astrophysics and Space Science Library}, Vol. 407,
  {Magnetic Fields in Diffuse Media}, ed. A.~{Lazarian}, E.~M. {de Gouveia Dal
  Pino}, \& C.~{Melioli}, 41

\bibitem[{{Heesen} {et~al.}(2009){Heesen}, {Beck}, {Krause}, \&
  {Dettmar}}]{Heesen}
{Heesen}, V., {Beck}, R., {Krause}, M., \& {Dettmar}, R.-J. 2009, \aap, 494,
  563

\bibitem[{{Hummel} \& {Saikia}(1991)}]{Hummel}
{Hummel}, E., \& {Saikia}, D.~J. 1991, \aap, 249, 43

\bibitem[{{Hummel} {et~al.}(1983){Hummel}, {van Gorkom}, \&
  {Kotanyi}}]{Hummel_1983}
{Hummel}, E., {van Gorkom}, J.~H., \& {Kotanyi}, C.~G. 1983, \apjl, 267, L5

\bibitem[{{Irwin} {et~al.}(2012){Irwin}, {Beck}, {Benjamin}, {Dettmar},
  {English}, {Heald}, {Henriksen}, {Johnson}, {Krause}, {Li}, {Miskolczi},
  {Mora}, {Murphy}, {Oosterloo}, {Porter}, {Rand}, {Saikia}, {Schmidt},
  {Strong}, {Walterbos}, {Wang}, \& {Wiegert}}]{Irwin}
{Irwin}, J., {Beck}, R., {Benjamin}, R.~A., {et~al.} 2012, \aj, 144, 43

\bibitem[{{Iwasawa} {et~al.}(2003){Iwasawa}, {Wilson}, {Fabian}, \&
  {Young}}]{Iwasawa}
{Iwasawa}, K., {Wilson}, A.~S., {Fabian}, A.~C., \& {Young}, A.~J. 2003,
  \mnras, 345, 369

\bibitem[{{Kellermann} {et~al.}(2007){Kellermann}, {Kovalev}, {Lister},
  {Homan}, {Kadler}, {Cohen}, {Ros}, {Zensus}, {Vermeulen}, {Aller}, \&
  {Aller}}]{Kellerman}
{Kellermann}, K.~I., {Kovalev}, Y.~Y., {Lister}, M.~L., {et~al.} 2007, \apss,
  311, 231

\bibitem[{{Krause}(2011)}]{Krause_2011}
{Krause}, M. 2011, ArXiv e-prints, arXiv:1111.7081

\bibitem[{{Kukula} {et~al.}(1995){Kukula}, {Pedlar}, {Baum}, \&
  {O'Dea}}]{Kukula}
{Kukula}, M.~J., {Pedlar}, A., {Baum}, S.~A., \& {O'Dea}, C.~P. 1995, \mnras,
  276, 1262

\bibitem[{{Kuo} {et~al.}(2011){Kuo}, {Braatz}, {Condon}, {Impellizzeri}, {Lo},
  {Zaw}, {Schenker}, {Henkel}, {Reid}, \& {Greene}}]{Kuo}
{Kuo}, C.~Y., {Braatz}, J.~A., {Condon}, J.~J., {et~al.} 2011, \apj, 727, 20

\bibitem[{{Li} {et~al.}(2008){Li}, {Li}, {Wang}, {Irwin}, \& {Rossa}}]{Li_2008}
{Li}, J.-T., {Li}, Z., {Wang}, Q.~D., {Irwin}, J.~A., \& {Rossa}, J. 2008,
  \mnras, 390, 59

\bibitem[{{Oosterloo} \& {van Gorkom}(2005)}]{Oosterloo}
{Oosterloo}, T., \& {van Gorkom}, J. 2005, \aap, 437, L19

\bibitem[{{Pappalardo} {et~al.}(2010){Pappalardo}, {Lan\c{c}on}, {Vollmer},
  {Ocvirk}, {Boissier}, \& {Boselli}}]{Pappalardo}
{Pappalardo}, C., {Lan\c{c}on}, A., {Vollmer}, B., {et~al.} 2010, \aap, 514,
  A33

\bibitem[{{Pearson} \& {Zensus}(1987)}]{Pearson}
{Pearson}, T.~J., \& {Zensus}, J.~A. 1987, in {Superluminal Radio Sources}, ed.
  J.~A. {Zensus} \& T.~J. {Pearson}, 1--11

\bibitem[{{Planck Collaboration} {et~al.}(2015){Planck Collaboration}, {Ade},
  {Aghanim}, {Arnaud}, {Ashdown}, {Aumont}, {Baccigalupi}, {Banday},
  {Barreiro}, {Bartolo}, \& et~al.}]{Planck_2015}
{Planck Collaboration}, {Ade}, P.~A.~R., {Aghanim}, N., {et~al.} 2015, ArXiv
  e-prints, arXiv:1511.05156

\bibitem[{{Ranalli} {et~al.}(2008){Ranalli}, {Comastri}, {Origlia}, \&
  {Maiolino}}]{Ranalli_2008}
{Ranalli}, P., {Comastri}, A., {Origlia}, L., \& {Maiolino}, R. 2008, \mnras,
  386, 1464

\bibitem[{{Roediger} \& {Br{\"u}ggen}(2008)}]{Roediger_2008}
{Roediger}, E., \& {Br{\"u}ggen}, M. 2008, \mnras, 388, 465

\bibitem[{{Roediger} {et~al.}(2006){Roediger}, {Br{\"u}ggen}, \&
  {Hoeft}}]{Roediger_2006}
{Roediger}, E., {Br{\"u}ggen}, M., \& {Hoeft}, M. 2006, \mnras, 371, 609

\bibitem[{{Soida} {et~al.}(2011){Soida}, {Krause}, {Dettmar}, \&
  {Urbanik}}]{Soida_2011}
{Soida}, M., {Krause}, M., {Dettmar}, R.-J., \& {Urbanik}, M. 2011, \aap, 531,
  A127

\bibitem[{{Sokoloff} {et~al.}(1998){Sokoloff}, {Bykov}, {Shukurov},
  {Berkhuijsen}, {Beck}, \& {Poezd}}]{Sokoloff_1998}
{Sokoloff}, D.~D., {Bykov}, A.~A., {Shukurov}, A., {et~al.} 1998, \mnras, 299,
  189

\bibitem[{{Strickland} {et~al.}(2002){Strickland}, {Heckman}, {Weaver},
  {Hoopes}, \& {Dahlem}}]{Strickland_2002}
{Strickland}, D.~K., {Heckman}, T.~M., {Weaver}, K.~A., {Hoopes}, C.~G., \&
  {Dahlem}, M. 2002, \apj, 568, 689

\bibitem[{{Strickland} {et~al.}(1997){Strickland}, {Ponman}, \&
  {Stevens}}]{Strickland_1997}
{Strickland}, D.~K., {Ponman}, T.~J., \& {Stevens}, I.~R. 1997, \aap, 320, 378

\bibitem[{{Tonnesen} \& {Bryan}(2009)}]{Tonnesen_2009}
{Tonnesen}, S., \& {Bryan}, G.~L. 2009, \apj, 694, 789

\bibitem[{{Urban} {et~al.}(2011){Urban}, {Werner}, {Simionescu}, {Allen}, \&
  {B{\"o}hringer}}]{Urban}
{Urban}, O., {Werner}, N., {Simionescu}, A., {Allen}, S.~W., \&
  {B{\"o}hringer}, H. 2011, \mnras, 414, 2101

\bibitem[{{Veilleux} {et~al.}(1999){Veilleux}, {Bland-Hawthorn}, {Cecil},
  {Tully}, \& {Miller}}]{Veilleux}
{Veilleux}, S., {Bland-Hawthorn}, J., {Cecil}, G., {Tully}, R.~B., \& {Miller},
  S.~T. 1999, \apj, 520, 111

\bibitem[{{Vikhlinin} {et~al.}(2001){Vikhlinin}, {Markevitch}, \&
  {Murray}}]{Vikhlinin_2001}
{Vikhlinin}, A., {Markevitch}, M., \& {Murray}, S.~S. 2001, \apj, 551, 160

\bibitem[{{Vollmer}(2009)}]{Vollmer_2009}
{Vollmer}, B. 2009, \aap, 502, 427

\bibitem[{{Vollmer} {et~al.}(2001){Vollmer}, {Cayatte}, {Balkowski}, \&
  {Duschl}}]{Vollmer_2001}
{Vollmer}, B., {Cayatte}, V., {Balkowski}, C., \& {Duschl}, W.~J. 2001, \apj,
  561, 708

\bibitem[{{Vollmer} \& {Huchtmeier}(2003)}]{Vollmer_2003}
{Vollmer}, B., \& {Huchtmeier}, W. 2003, \aap, 406, 427

\bibitem[{{Vollmer} {et~al.}(2013){Vollmer}, {Soida}, {Beck}, {Chung},
  {Urbanik}, {Chy{\.z}y}, {Otmianowska-Mazur}, \& {Kenney}}]{Vollmer_2013}
{Vollmer}, B., {Soida}, M., {Beck}, R., {et~al.} 2013, \aap, 553, A116

\bibitem[{{Vollmer} {et~al.}(2007){Vollmer}, {Soida}, {Beck}, {Urbanik},
  {Chy{\.z}y}, {Otmianowska-Mazur}, {Kenney}, \& {van Gorkom}}]{Vollmer_2007}
---. 2007, \aap, 464, L37

\bibitem[{{Vollmer} {et~al.}(2010){Vollmer}, {Soida}, {Chung}, {Beck},
  {Urbanik}, {Chy{\.z}y}, {Otmianowska-Mazur}, \& {van Gorkom}}]{Vollmer_2010}
{Vollmer}, B., {Soida}, M., {Chung}, A., {et~al.} 2010, \aap, 512, A36

\bibitem[{{Vollmer} {et~al.}(2008){Vollmer}, {Soida}, {Chung}, {van Gorkom},
  {Otmianowska-Mazur}, {Beck}, {Urbanik}, \& {Kenney}}]{Vollmer_2008}
---. 2008, \aap, 483, 89

\bibitem[{{We{\.z}gowiec} {et~al.}(2012){We{\.z}gowiec}, {Urbanik}, {Beck},
  {Chy{\.z}y}, \& {Soida}}]{Wezgowiec}
{We{\.z}gowiec}, M., {Urbanik}, M., {Beck}, R., {Chy{\.z}y}, K.~T., \& {Soida},
  M. 2012, \aap, 545, A69

\bibitem[{{We{\.z}gowiec} {et~al.}(2011){We{\.z}gowiec}, {Vollmer}, {Ehle},
  {Dettmar}, {Bomans}, {Chy{\.z}y}, {Urbanik}, \& {Soida}}]{Wezgowiec_Mach}
{We{\.z}gowiec}, M., {Vollmer}, B., {Ehle}, M., {et~al.} 2011, \aap, 531, A44

\bibitem[{{Wiegert} {et~al.}(2015){Wiegert}, {Irwin}, {Miskolczi}, {Schmidt},
  {Mora}, {Damas-Segovia}, {Stein}, {English}, {Rand}, {Santistevan},
  {Walterbos}, {Krause}, {Beck}, {Dettmar}, {Kepley}, {Wezgowiec}, {Wang},
  {Heald}, {Li}, {MacGregor}, {Johnson}, {Strong}, {DeSouza}, \&
  {Porter}}]{Wiegert_2015}
{Wiegert}, T., {Irwin}, J., {Miskolczi}, A., {et~al.} 2015, \aj, 150, 81

\bibitem[{{Yoshida} {et~al.}(2002){Yoshida}, {Yagi}, {Okamura}, {Aoki},
  {Ohyama}, {Komiyama}, {Yasuda}, {Iye}, {Kashikawa}, {Doi}, {Furusawa},
  {Hamabe}, {Kimura}, {Miyazaki}, {Miyazaki}, {Nakata}, {Ouchi}, {Sekiguchi},
  {Shimasaku}, \& {Ohtani}}]{Yoshida_2002}
{Yoshida}, M., {Yagi}, M., {Okamura}, S., {et~al.} 2002, \apj, 567, 118

\bibitem[{{Yoshida} {et~al.}(2004){Yoshida}, {Ohyama}, {Iye}, {Aoki},
  {Kashikawa}, {Sasaki}, {Shimasaku}, {Yagi}, {Okamura}, {Doi}, {Furusawa},
  {Hamabe}, {Kimura}, {Komiyama}, {Miyazaki}, {Miyazaki}, {Nakata}, {Ouchi},
  {Sekiguchi}, \& {Yasuda}}]{Yoshida_2004}
{Yoshida}, M., {Ohyama}, Y., {Iye}, M., {et~al.} 2004, \aj, 127, 90

\bibitem[{{Zhuravleva} {et~al.}(2013){Zhuravleva}, {Churazov}, {Kravtsov},
  {Lau}, {Nagai}, \& {Sunyaev}}]{Zhuravleva_2013}
{Zhuravleva}, I., {Churazov}, E., {Kravtsov}, A., {et~al.} 2013, \mnras, 428,
  3274

\end{thebibliography}

\clearpage

\end{document}